\RequirePackage{fixltx2e}
\documentclass[english,review,1p,times]{IEEEtran}
\usepackage{amsmath}
\usepackage{amsthm}
\usepackage{newtxmath}
\usepackage[T1]{fontenc}
\usepackage[latin9]{inputenc}
\usepackage{mathtools}
\usepackage{graphicx}
\usepackage{setspace}
\makeatletter
\numberwithin{equation}{section}
\numberwithin{figure}{section}
\numberwithin{table}{section}
\theoremstyle{plain}
\newtheorem{thm}{\protect\theoremname}
\theoremstyle{remark}
\newtheorem{rem}[thm]{\protect\remarkname}
\theoremstyle{definition}
\newtheorem{defn}[thm]{\protect\definitionname}
\theoremstyle{plain}
\newtheorem{lem}[thm]{\protect\lemmaname}
\theoremstyle{plain}
\newtheorem{cor}[thm]{\protect\corollaryname}
\theoremstyle{remark}
\newtheorem{claim}[thm]{\protect\claimname}

\usepackage[american]{babel}

\makeatother

\usepackage[style=authoryear,backend=bibtex]{biblatex}
\providecommand{\claimname}{Claim}
\providecommand{\corollaryname}{Corollary}
\providecommand{\definitionname}{Definition}
\providecommand{\lemmaname}{Lemma}
\providecommand{\remarkname}{Remark}
\providecommand{\theoremname}{Theorem}

\begin{document}
\title{Algebra of N-event synchronization}
\author{ \emph{Ernesto Gomez}(1), \emph{Keith E. Schubert}(2), and \emph{Khalil Dajani}(1)

(1) California State University San Bernardino, School of Computer Science and Engineering, (2) Baylor University, Department of Electrical and Computer Engineering}
We have previously (\textcite{algebra2011})
defined synchronization as a relation between the times at which a
pair of events can happen, and introduced an algebra that covers all
possible relations for such pairs. In this work we introduce the synchronization
matrix, to make it easier to calculate the properties and results
of $N$ event synchronizations, such as are commonly encountered in
parallel execution of multiple processes. The synchronization matrix
leads to the definition of N-event synchronization algebras as specific
extensions to the original algebra. We derive general properties of
such synchronization, and we are able to analyze effects of synchronization
on the phase space of parallel execution introduced in \textcite{entropy2016}.


\maketitle
$\maketitle$

\section{\label{sec:Current-work}Algebra of synchronization}

Real time advances monotonically; to achieve a particular ordering
between events \emph{a} and \emph{b}, it is sufficient to block \emph{a}
or \emph{b} or both and release them when the relation is satisfied.
Our Boolean algebra of synchronization ~\textcite{algebra2011,GS10}
describes all possible relations between the times associated with
any number of event pairs. In our previous work, we described how
a set of synchronizations can imply relations that are not explicit
in the code, leading to unanticipated results such as deadlock. At
that time we handled N-event synchronizations from a graph of connections.
In this work we give systematic methods to describe and analyze any
number of events. These methods become more relevant after \textcite{entropy2016}.
In that work, we showed a phase space for parallel execution and its
relation to execution entropy, N-event analysis is required to predict
the effect of synchronization on execution phase space and estimate
the work required.

\subsection{Algebra of synchronization for event pairs\label{subsec:Algebra-of-synchronization} }

Synchronization is often described in the context of operating systems
(see \textcite{Tan97,Tan01,Stal04,Sil05}, ) usually through examples
of synchronization mechanisms. The literature on parallel computing
such as ~\textcite{Fos94,Jor03,Scott05} has mostly also treated
synchronization through specific example. ~\textcite{AptOld97} establish
a more theoretical approach to synchronization between parallel processes,
using the semantics of process waiting. However they also do not give
a general definition of synchronization.

In ~\textcite{algebra2011}\parencite*{GS10} we introduced a formal
definition of synchronization as a relation between sets of allowable
times for two events. This led to a Boolean algebra of synchronization
based on subset relations between event pairs. It follows that a synchronization
between multiple events can be described in terms of a collection
of event pairs, and we was used it in that paper to prove some properties
of semaphores and deadlock. 

We associate each event with a real number denoting the time at which
it occurs. There are precisely 6 possible relations between ordered
pairs of numbers - adding $1$ for union of all relations and 0 for
the empty set, we have the set of binary relations , together with
operators union $\cup,$intersection $\cap$, and complement $\sim$
we have the algebra of synchronization. The algebra is complete for
event pairs, since there are no other mathematical relations between
two numbers.

\begin{figure*}[t]
\caption{\label{fig:Algebra-of-binary}Algebra of binary synchronization ~\textcite{algebra2011,GS10}}
\includegraphics[width=3in]{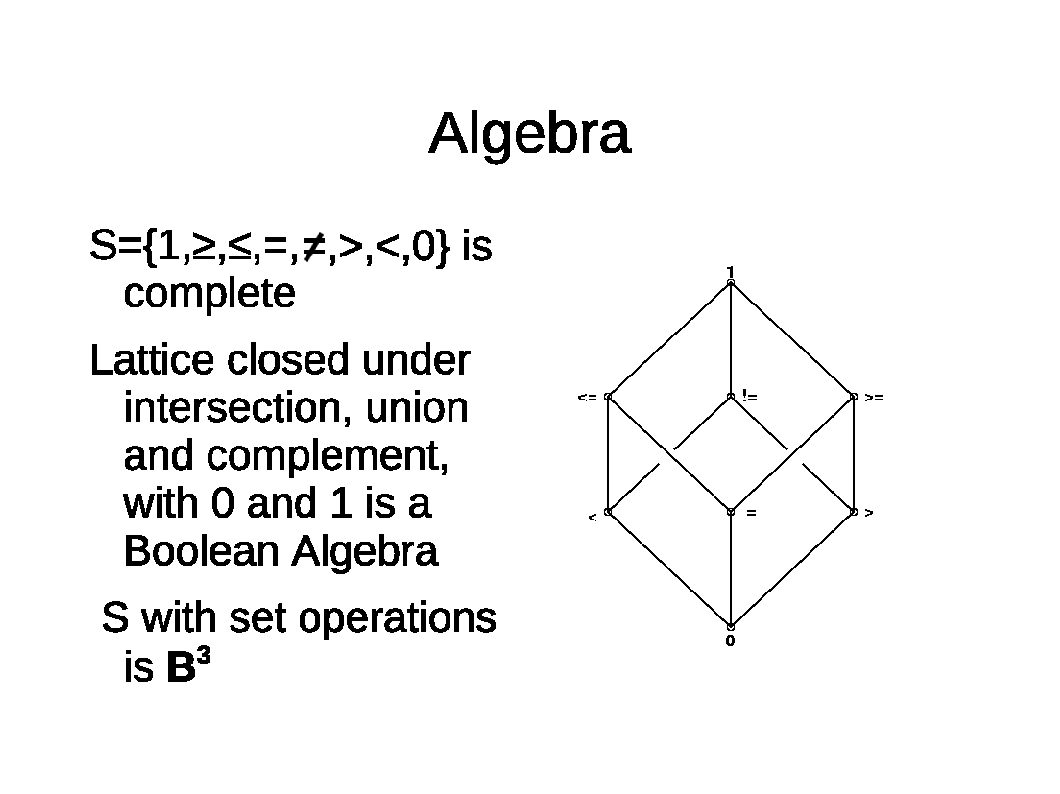} 
\end{figure*}

Let $s\in S$, then the ordered tuple $s(A,B)$ is the set of all
pairs of times $A$ and $B$ that satisfy $AsB$. In figure\ref{fig:Algebra-of-binary}
we show how $S$ is displayed as a graph, with $1$ and $0$ at opposite
vertices. That graph is a lattice (see ~\textcite{GILL76},~\textcite{BML99}),
each node of the graph except TOP$=1$ and BOTTOM=$0$ is on multiple
paths from TOP to BOTTOM, and it is easily verified that, for any
pair of nodes on the same path, the node closer to TOP is a superset
of the node closer to BOTTOM. It is simple to verify that the set
$\{1,\ge,\le,=,\ne,>,<,0\}$ including all the nodes in the graph,
when interpreted as labels denoting sets of ordered pairs, is closed
under union, intersection and complement, and is therefore an algebra
of subsets. Immediately this means that $S=<\{1,\ge,\le,\ne,>,<,0\},\cup,\cap,\sim>$
(the set of relations with operations union, intersection and complement)
is a Boolean algebra, isomorphic to $B^{3}$ because it has $2^{3}$elements.

The lattice for a Boolean algebra $B^{n}$ is an n-dimensional cube
or hypercube (see ~\textcite{GILL76}). See figure \ref{fig:Algebra-of-binary}

As a Boolean algebra, the entire body of rules developed for such
algebras (for example DeMorgan's law) is immediately applicable to
$S$.

The elements of $S$ represent the relation between two sets of points
- for instance $A>B$ means that a selected point in set $A$ is greater
than every point in set $B$; so $B$ is bounded from above; correspondingly
any point we select in $B$ is less than every point in $A$. $A=B$
means that the range of both sets is the same, it is always possible
to select matching points in $B$ and $A$. A further implication:
Since $S$ is an algebra of pairs ordered by relations between numbers,
we have some relations that are antisymmetric - \{$\ge,\le.>,<\}$
so we add a mirror $\mu$ unary operator. It is immediate that $S$
is closed under $\mu$, and $\mu(a\{\cap|\cup|\sim\}b)$$\Leftrightarrow\text{\ensuremath{\mu(a)\{\cup|\cap|\sim\}\mu(b)} }$-
that is, $\mu$ is associative with the standard Boolean operators.

Summarizing, we have:

$s\in\{1,\ge,\le,\ne,>,<,0\},$

$S$ is the set of number pairs $(t_{1},t_{2})$ where $t_{1}$ $s$
$t_{2}$ is the relation between times $t_{i}$. 

$S$ is an algebra over the set of ordered pairs of numbers, complete
and closed under the given operations (union, intersection, complement
and mirror), is Boolean algebra of subsets isomorphic to $B^{3}$
(see ~\textcite{algebra2011,GS10}). When considering >2 events,
we will refer to this algebra as $S_{2},$and its extension to $n$
events as $S_{n}$ 

We have:
\begin{itemize}
\item $\cup,\cap$ (union, intersection) are associative and commutative
over each other, by properties of subset algebras. 
\item $\sim,\mu$ (complement, mirror) are commutative (\emph{complement}
is the same relation as \emph{not})
\item $\sim,\mu$ are associative over $\cap,\cup$ 
\item deMorgan's laws: $\sim(A\cup B)=\sim A\cap\sim B$ and $\sim(A\cap B)=\sim A\cup\sim B$ 
\end{itemize}
We will use the symbols in $S$ as label for a synchronization defined
by the given symbol, as well as in numeric expressions.

Consider now the effects of synchronization on a single event. It
is evident from $S$ that a single event can be bounded from above,
below, above and below or unbounded. This gives a Boolean algebra
of boundedness as a sub-algebra of $S$ isomorphic to $B^{2}$, this
is described in ~\textcite{algebra2011,GS10} and furthered detailed
here. Formal development of N-event algebra, the synchronization matrix
and other properties is developed below.

\subsubsection{Sub-algebras of $S_{2}$ }

We identify 2 sub-algebras, $\text{\text{\ensuremath{L}}}_{1}{1,\ge,\le,=}$,
$\text{\ensuremath{L}}_{0}={\ne,>,<,0}$. $L_{1}$ includes the =
relation, $L_{0}$ does not. Every relation in$L_{1}$ includes points
on the boundary between sets, whereas $L_{0}$ does not include points
on the boundary. We can view $L_{1}$ as expressing relations between
closed sets, and $L_{0}$ as relations between non-overlapping open
sets. Inmeadiately we have that $\cup,\cap$ are closed in each of
the sub-algebras, as is mirror $\mu$ which just reverses the order.
The $\sim$ (not) relation from $S_{2}$ shifts between $L_{1}$ and
$L_{0}$ since $\sim\ne$ is the same as ; we take complement to be
the same as mirror in the sib-algebras. (see \textcite{GS10}.

\subsection{Properties of the binary synchronization algebra }

\subsubsection{\label{sec:Algebra-of-bounded}Algebra of bounded event sets $\beta$}

In order to deal with synchronizations involving multiple events,
we need to account for cases in which a single event synchronizes
with more than one other event. We describe how synchronization imposes
boundaries on events to analyze this. Since in $S_{2}$ is an ordered
pair of events $(t_{1},t_{2})$, where each $t_{i}$ is a set of numbers
corresponding to allowed event times, synchronization imposes a boundary
on membership in each set. We use notation LH and RH to denote sets
textually to the left and to the right of a synchronization operator,
respectively: 
\begin{enumerate}
\item No boundaries: $1,\ne$ do not impose any upper or lower bounds on
sets; in either case the range of allowed values extends to positive
and negative infinity. $\ne$ excludes a point (or continuous subset
of points) corresponding to an event in RH from the set of points
in LH (and vice-versa). The excluded point(s) are bounded above and
below, but do not impose upper or lower bounds on sets RH and LH. 
\item Bounded above and below: $=,0$ impose upper and lower bounds on both
LH and RH. The bounds on the LH set are the same as on RH, in the
case of $0$ synchronization, upper and lower bounds may coincide
allowing no points in either set, but it is possible for two events
to bind a third - $a>x>b$ or $a\ge x\ge b$. The first case yields
0, but does not have to be empty if $a>b$, this makes $x$ an open
set not including its boundaries. The second case yields $=$ , the
set $x$ includes boundaries. A mixed case is still $0$ since the
relation between $a$ and \textbf{$b$} satisfies $a\cap b=\phi$. 
\item Bounded above or below: $\ge,>$ or $\le,<$impose a single upper
(lower) bound, differing only in that $\ge,\le$ include the boundary
value in both RH and LH sets, and $>,<$ exclude the boundary from
both. 
\end{enumerate}
We get 8 cases, 4 isomorphic to $L_{0}$: $B_{0}$$\Leftrightarrow$$L_{0}$
(excluding boundaries) and 4 isomorphic to $L_{1}$: $B_{1}$$\Leftrightarrow$$L_{1}$
(including boundaries). Their combination gives the full algebra $\text{\ensuremath{\beta}}\Leftrightarrow S_{2}$.
As before, the union of an element in $B_{0}$ with an element in
$B_{1}$ yields an element in $B_{1}$, and intersection yields an
element in $B_{0}$. $\beta$ is complete; it expresses all possible
boundaries that can be imposed on an event by a synchronization with
one other event. We now consider boundaries imposed by multiple events. 
\begin{rem}
Notation: Since algebras $\beta$, $B_{0}$, $B_{1}$ are isomorphic
to $S$, $L_{0}$, $L_{1}$ it is convenient to use the same notation.
Given sets $t_{1},t_{2}$ representing events, with a synchronization
$s_{12}$, then the set $t_{1}$ has boundary $s_{12}$ with respect
to itself, and $t_{2}$ has boundary $s_{21}=\mu(s_{12})$. For example,
if $s_{12}$ is $>$, then $t_{1}$ is bounded below by $t_{2}$ and
the boundary is $>$ , the same label as the synchronization. 
\end{rem}
\begin{thm}
\label{thm:n-events-from2}Every synchronization between $n>2$ events
can be described using binary synchronizations in the algebra $S_{2}$
between the given events 
\end{thm}
\begin{IEEEproof}
Base case: events $e_{1,},e_{2}$ with synchronization $s_{12}\in S_{2}$,
completely defines the relation between the events (proved in ~\textcite{algebra2011}).

Induction: Given $n$ events, add event $n+1$ , then $\forall_{i<0\le n}$
add pairs $(e_{i},e_{n+1})$ with synchronization $s._{i,n+1}$ ,
which completely defines the relation between the added event and
each other event. Since relations between every pair in $n$ are completely
described in $S_{2}$, and this is also true for every relation between
the original $n$ events and event $n+1$ is completely described
in $S_{2}$, then every relation between a set of $n+1$ events is
described in $S_{2}$. 
\end{IEEEproof}

\section{\label{sec:2-event-matrix}Synchronization matrix and extension to
N events}

We introduce a matrix notation that makes it easier to describe synchronizations
of more than 2 events. With > 2 events, we can get implied synchronizations
- for example consider 3 events such that: $e_{1}>e_{2}$ and $e_{2}>e_{3}$.
It follows that $e_{1}>e_{3}$ which is not coded, but is implied.
Implied synchronizations happen because we have transitivity for some
synchronizations. We will describe transitivity, and show an algorithm
that uses it to compute closure of a synchronization matrix, which
includes all the implied synchronizations and gives us the boundedness
condition on each event.

\subsection{\label{subsec:s-matrix} Synchronization matrix}
\begin{defn}
\textbf{\textit{Binary Synchronization Matrix}}: $\begin{bmatrix}s_{11} & \text{\ensuremath{s_{12}}}\\
\text{\ensuremath{s_{21}}} & s_{22}
\end{bmatrix}$. The $s_{ij}$ are in the synchronization algebra $S_{2}$, $s_{ji}=\mu(s_{ij})$
and diagonal elements $s_{ii}$ are defined as 1 since they relate
an event to itself. 
\end{defn}
Following are the synchronization matrices for algebra $S_{2}$

$M_{2}$=\{$\begin{bmatrix}1 & 1\\
1 & 1
\end{bmatrix}$,$\begin{bmatrix}1 & \ge\\
\le & 1
\end{bmatrix}$, $\begin{bmatrix}1 & \le\\
\ge & 1
\end{bmatrix}$, $\begin{bmatrix}1 & =\\
= & 1
\end{bmatrix}$, $\begin{bmatrix}1 & \ne\\
\ne & 1
\end{bmatrix}$$,$$\begin{bmatrix}1 & >\\
< & 1
\end{bmatrix}$, $\begin{bmatrix}1 & <\\
> & 1
\end{bmatrix}$,$\begin{bmatrix}1 & 0\\
0 & 1
\end{bmatrix}\}$.

The upper right element $s_{12}$ is the synchronization between ordered
event pair (1,2). The lower left $s_{21}$ represents the event pair
(2,1) which reverses (1,2), so the corresponding synchronization is
the mirror, $s_{21}=\mu(s_{12})$ . Each matrix is fully determined
by the element above the diagonal, so $M_{2}$ has the same relations
as $S_{2}$.
\begin{rem}
\emph{It would be convenient to set the elements $s_{ii}$ to their
boundedness in $\beta$ . We do not do this; the resulting matrices
in $M_{n},n>2$ would not be closed under $\cap,\cup$ because diagonal
element boundedness is always defined by intersection, but the off-diagonal
elements $s_{ij}$would be subject to $\cap$ or $\cup$. Assume matrices
A and B with diagonal elements in $\beta$. For 3 events, we have:
$a_{ii}=a_{ij}\cap a_{ik}$, $b_{ii}=b_{ij}\cap b_{ik}$. Closure
implies $a_{ii}\cup b_{ii}=$$(a_{ij}\cap a_{ik})\cup(b_{ij}\cap b_{ik})$
and $a_{ii}\cap b_{ii}=$$(a_{ij}\cap a_{ik})\cap(b_{ij}\cap b_{ik})$,
but these four statements taken together would violate deMorgan's
laws.}
\end{rem}
We now define element-wise operations on the matrix:
\begin{defn}
\label{def:operations-on-M2}Operations on $M_{2}$ follow from the
operations on $S_{2}$; given $m_{i},m_{j},\in M_{2}$ with o$\in$
Op=$\{\cap,\cup,\sim,\mu$\} applied on each element of a matrix $M\in M_{2}$: 
\end{defn}
\begin{itemize}
\item If o$\in$Op is unary, then $o[M]\Leftrightarrow\forall_{i,j}o(m_{i,j}$) 
\item If o$\in$Op is binary, then $o[p,q]\Leftrightarrow\forall_{i,j}o(p_{i,j},q_{i,j})$ 
\end{itemize}
We show $M_{2}$ is closed under these operations: 
\begin{lem}
\label{lem:M2 is closed}The set $M_{2}$ is closed under Op= $o\in$$\{\cap,\cup,\sim,\mu\}$ 
\end{lem}
\begin{IEEEproof}
Binary operators on matrices act on individual elements, and are closed
in $S$. Mirror ($\mu)$ symmetry is preserved across the diagonal
because synchronization matrices all have $s_{ij}=\mu(s_{ji})$ and
$\mu(aob)$=$\mu(a)o\mu(b)$ for $o\in S$ (associative property) 
\end{IEEEproof}

\subsection{Synchronization matrix for n>2 }

An $n$ event synchronization is specified by the set of all synchronizations
between event pairs (theorem \ref{thm:n-events-from2}). This can
be represented as a synchronization matrix. We now construct an algebra
$S_{n}$ for $n$ events, with elements $s_{ij}$ denoting relations: 
\begin{defn}
\label{def:n-element-synchronization-matrix}\textbf{n-element synchronization
matrix}. Using the basic layout of the binary synchronization matrix
\ref{sec:2-event-matrix}: Number the events $i,j\in(1...n)$, this
gives us $n^{2}$ ordered pairs of events $(i,j)$ . For an $nxn$
matrix $M,$ we map each event pair to element $m_{ij}\Leftrightarrow(i,j)$.
Set the diagonal elements $m_{ii}=1$. 
\end{defn}
We are left with $(n^{2}-n)$ off-diagonal event pairs. However, every
element $s_{ij}$ above the diagonal determines a corresponding element
$s_{ji}=\mu(s_{ij})$ below the diagonal, so only half the event pairs
are independent. Therefore 
\begin{lem}
\label{lem:The--elements}The $p=(n^{2}-n)/2$ elements above the
diagonal fully determine the matrix. Since each of the entries above
the diagonal must be an element of $S_{2}$ , there are precisely
$8^{p}=2^{3p}$ different $nxn$ synchronization matrices of $n$
events. 
\end{lem}
\begin{cor}
The algebra of synchronization $S_{n}$ is isomorphic to Boolean $B^{3p}$,
since it is a complete algebra of subsets with $2^{3p}$ elements. 
\end{cor}

\subsection{\label{subsec:Atoms}Atoms }

A Boolean algebra $B^{n}$ has $2^{n}$ elements, represented graphically
as an n-dimensional hypercube of side 1. As $n$ increases, it becomes
impractical to list all the elements of the algebra, but we can find
$n$ elements $a_{i}$ which we call \emph{atoms,} such that $a_{i}\in B^{n}$
and $a_{i}\cap a_{j}=0$ for $i\ne j$. With $\cup$ as addition,
the $n$ atoms are an additive basis for $B^{n}$ (~\textcite{BML99},\textcite{JAC51},~\textcite{GILL76})
For example, $S_{2}$ is isomorphic to $B^{3}$, and its $3$ atoms
are $\{=,<,>\}$.

We visualize the $n$ atoms in the lattice hypercube as the $n$ arcs
$(0,i)$, equivalent to $n$ directions in a Cartesian coordinate
system. It is evident that we can reach any point by in the hypercube
from any other by taking $\le n$ steps along arcs parallel to the
atoms, equivalent to $\cup$ of $n$ atoms

To construct the synchronization matrices for the $3p$ atoms of $S_{n}:$
Take the $0$ matrix in $S_{n}$. (diagonal elements are $1$, off-diagonal
elements are $0$). Each above diagonal element $m_{i,j}$ with $i<j$
gives 3 atoms, by setting it to each of the 3 atoms of $S_{2}$, set
below-diagonal $m_{j,i}$to $\mu(m_{i,j})$. Repeat for each element
above the diagonal.
\begin{itemize}
\item \label{sec:Calculation-of-boundedness}Calculation of boundedness
\end{itemize}
We have defined an algebra of bounds on events in section \ref{sec:Algebra-of-bounded}.
Since these apply to single events, we have no need to extend them
for synchronization of more events. However, the bounds on any specific
event may include restrictions of implied synchronizations, so we
need to take this into account.

We compute the boundedness on event $i$ as follows:
\begin{thm}
\label{lem:The-boundedness-condition}The boundedness condition on
an event $i$ in a synchronization matrix is $1\cap_{i\ne j}m_{ij}$. 
\end{thm}
\begin{IEEEproof}
Immediate for $s\in S_{2}$ (see \ref{sec:2-event-matrix}) because
$1\cap s=s,$ every $m_{ij}$ is an additional (and) restriction on
the bounds of event $i$, $m_{ij}$ in row $i$ includes all pairs
$(i,j)$ 
\end{IEEEproof}
\begin{cor}
\label{cor:The-boundedness-of}The boundedness of an event in a synchronization
matrix is in the boundedness algebra $\beta$ (see \ref{sec:Algebra-of-bounded}). 
\end{cor}
\begin{IEEEproof}
Let $p$ be the number of elements above the diagonal (see lemma \ref{lem:The--elements})
, the set of $2^{3p}$ (see lemma \ref{lem:The--elements}) $n$ element
synchronization matrices (definition \ref{def:n-element-synchronization-matrix})
forms the algebra of synchronization $S_{n}$ for $n$ events. Since
$S_{n}$is a closed and complete algebra of subsets, it is Boolean. 

The algebras of synchronization $S_{2}$ and boundedness $\beta$
are closed under binary operators $\cap,\cup$ and unary operators
$\sim,\mu$ then applying these element-wise to NxN synchronization
matrices yields another NxN matrix. This matrix is a synchronization
matrix because $S_{2}$ is Boolean and closed under $\sim,\mu$ and
so preserve the symmetry between upper and lower diagonal areas.

By theorem \ref{thm:n-events-from2} we have that an n-event synchronization
is composed of event pairs.

By lemma \ref{lem:The-boundedness-condition} we have that the diagonal
elements are determined by the synchronizations in the same row.

From section \ref{sec:2-event-matrix} we have that elements below
the diagonal are determined by mirror of above diagonal entries.

The $nxn$ matrix has $n^{2}$ entries, but only the elements above
the diagonal are independent. Subtracting $n$ diagonal elements and
dividing by 2 to get half the off-diagonals, we have $p=(n^{2}-n)/2$
independent elements in the synchronization matrix.

There are $2^{3}$ elements in algebra $S_{2};S_{n}$ with $p$ independent
elements gives $2^{3p}$ combinations of values

By induction on lemmas \ref{lem:M2 is closed} and \ref{lem:The--elements}
the $n$ element synchronization matrices are closed under set and
mirror operators, and the set of such matrices is complete because
the algebra $S$ (sec. \ref{subsec:Algebra-of-synchronization}) is
complete.
\end{IEEEproof}
We have that boundedness of each event in a matrix $S_{n}$ is in
$\beta$, so it is represented by a vector $B$ of size $n$, such
that each $b_{i}$ represents the boundedness on event $e_{ii}$ in
the synchronization matrix.

For a specific $e_{ii}$, the synchronizations that explicitly apply
are $s_{ij},j\ne i$ (row i) and $s_{ji},j\ne i$ (column i); since
column i is the $\mu$ of row i (by definition of a synchronization
matrix) we need only consider one of the two; we choose to use synchronizations
on the same row for readability. Synchronizations on row $i$ have
the form $s_{ij}$ imposing a restriction on event $j$. Therefore
the effect on $i$ is $\mu$($s_{ij})$ and the effect of all the
synchronizations in row i on $e_{ii}$ is $1\cap\prod_{j}\text{\ensuremath{\mu}(}s_{i\ne j})$,
where $\cap$ takes the expected meaning of applying all the restrictions
on the row to the same event. If we need an \emph{or} condition it
can be specified by union of synchronization matrices.

\subsection{\label{sec:Transitivity:}Transitivity:}

We need boundedness to define transitivity, because matching boundaries
transmit the effect of a synchronization. For example, $a>b>c$ transmits
the boundary imposed by > from $a$ to $c$. In $a>b<c$ the boundaries
established by > and < do not match and no relation is enforced between
a and c.

We distinguish the following cases, in order of evaluation: 
\begin{enumerate}
\item if $e_{ik},\in\{\ge,>\}$ ,$e_{kj}\in$ $\{\le,<\}$ - or mirror of
these : lower (upper if $\mu(e_{i,j})$ and $\mu(e_{k,j})$) bound
is transmitted, but does not enforce any relation between events on
either side, so $e_{ij}=1$. 
\item If $e_{ik},e_{kj}\in\{\ge,>\}$ (alternately both in $\{\le,<\}$
: $e_{ij}=e_{ik}\cap e_{kj}$. The boundary is transmitted, the edge
point is in the set only if included by both $e_{ik},e_{kj}.$ 
\item If either $e_{ik},e_{kj}$ is $=$ : upper and lower boundaries transmitted
and included, if $e_{ik}$ is $=$ then $e_{ij}=e_{kj}$, else if
$e_{kj}$ is = then $e_{ij}=e_{ik}$. 
\item If either $e_{ik},e_{kj}$ $\in\{1,\ne\}$ : no bounds are transmitted,
$e_{ij}=1$. 
\end{enumerate}

\section{Closure and semantics}

A synchronization matrix for $n$events may be defined using explicitly
declared synchronizations; these relations may imply other relations
not explicitly written into the matrix, including deadlock (if a relation
between different events resolves to 0). The closure of the synchronization
matrix computes all the relations enforced by the declared synchronizations,
and may be extended by writing the boundedness condition on each event
into the diagonal. Therefore it gives us the meaning of the synchronization.

Given: a a list of binary synchronizations $L$ between a set of $n$events.
To produce the closure of the synchronization matrix: 
\begin{enumerate}
\item Initialize every element $=1$ . 
\item Set every defined synchronization relation $s_{ij}$ by an appropriate
element $s\in S$, 
\item For every $s$ entered in step 2 between events $(i,j)$ $s_{ij}=s\cap s_{ij}$
and$s_{ji}=\mu(s_{ij})\cap s_{ji}$ (if either $s_{ij}$ or $s_{ji}$
was not initially 1, verify that $s_{ij}=\mu(s_{ji})$, if it is not,
replace $s_{ij}=\mu(s_{ji})\cap s_{ij}$ and check $s_{ji}$again) 
\item For every pair of indices $(i,j),i\ne j$ : calculate the transitive
value $t=s_{ik}s_{kj}$ for all $k\ne i,j$, and set $s_{ij}=t\cap s_{ij}$.
repeat until no change in any $s_{ij}$. By construction the result
is a synchronization matrix, because elements on the diagonal are
1, every element $(i,j),i\ne j$ is set to a synchronization in $S$,
and every $s_{ij}=\mu(s_{ji})$. 
\item Set $e_{ii}=1\cap$$\prod_{k\ne i}\mu(s_{ik})$ : each element in
trace is the effect of sync applied from the left onto the diagonal,
mirror the off-diagonal elements because index ij on row i denotes
the action of i on j and we want the action of j on i. 
\item If either $e_{ik},e_{kj}$ is 0 : we have an impossible condition
so $e_{ij}=0.$ 
\item The diagonal is computed as described in \ref{sec:Calculation-of-boundedness} 
\end{enumerate}
\begin{rem}
In calculating transitivity in step 4, we need to consider the implementation
of a critical section $\ne$, for example as a semaphore. The $\ne$
relation is not transitive; and the original semaphore definition
by \textcite{Dij65} does not deadlock, but in practice a queue is
frequently attached to hold processes that try to access the critical
section when it is busy. For example, if a process $p_{1}$ enters
a semaphore, a process $p_{2}$ that requests the same code is placed
in a queue. Rather than $p_{1}\ne p_{2}$ we actually get $p_{1}<p_{2}$.
We give a detailed discussion in \textcite{GS10}. We show that fairness
condition established by a queue allows semaphores to deadlock some
of the time. For practical computations of closure we may choose to
deal with $\ne$ as either > or < , depending on which alternative
is transitive, since the queue can lead to deadlock unlike the pure
$\ne$. We are able to establish the possibility of deadlock in the
closure, at the cost of displaying only a worst case of the synchronization.
\end{rem}
Although closure is a useful and compact description of a synchronization,
it does not retain all the properties of a synchronization matrix.
In the definition of a synchronization matrix the diagonal elements
are $1$, meaning that an event does not restrict itself. Boundedness
is not preserved by element-wise union and intersection operations
on the closure matrix (although operations are correct for the off-diagonal
elements). When combining closure matrices using the standard definitions
in the algebra, we need to recalculate the diagonal of the resulting
matrices.

To consider ``what does a synchronization mean?'' we need to move
past the relations that synchronization imposes to the events that
are ordered by them. The events themselves are actions that are dependent
in some way on time - for example, state changes, start or end of
a process, receipt or sending a signal, whatever. In general we can
assign names to these events for our convenience. Names of events,
even though they may be arbitrary, are not to be regarded as simple
interchangeable labels, they are added as an extra property of the
event (for example, the name $\pi$ references a particular real number,
we may refer to $\pi$ by another label or alias, but the name would
lose its usefulness if it could designate different numeric values).
We can label a set of $n$ events using numbers $(1\ldots n)$, or
any other set of symbols, and use these labels to reference a row
in a synchronization matrix, with the understanding that we can now
speak of changing the position of an event in a synchronization matrix
as a re-labeling.

The actual index label is not itself significant, but synchronizations
we specify on a row of a matrix are affected by the relation between
the index of two events. For example, if an event labeled A must happen
before B, then the time relation between them is A<B. However, if
event B is in a synchronization matrix at row $1$, and event A is
at row 2, then the synchronization $m_{12}$ describes A<B. Re-labeling
$m_{11}=A$and $m_{22}=B$ would reverse (operator $\mu$) the relation
between $m_{11}$ and $m_{22}$ while preserving the meaning A<B.
For $n>2$, such re-labeling may affect the synchronizations between
re-labeled events and all other events in the matrix. Function \textit{eventswap}
(specified in smat-u.sce) describes the changes in the synchronization
matrix that preserve the event order while changing the event indices:

\begin{figure}[h]
Code to switch event indices in synchronization matrix

\includegraphics[scale=0.5]{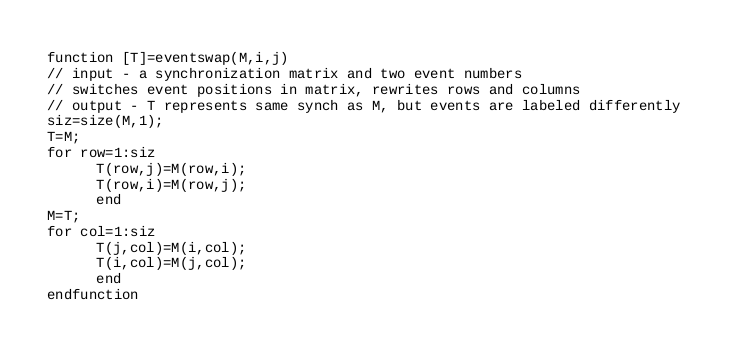}
\end{figure}

\begin{verbatim*}

\end{verbatim*}
Eventswap reads the original matrix and then writes the swapped rows
and columns into the copy. By definition of a synchronization matrix,
all synchronization relations on the same row/column involve the event
selected on the diagonal. Therefore order of operations in eventswap
does not matter. 
\begin{claim}
Given a synchronization matrix representing $N$ events and their
relations, any permutation performed by a sequence of eventswap actions
represents the same synchronization since it preserves the relations
between the actual events. From this we conclude that any permutation
of events performed through a succession of eventswap preserves the
relations between actual events and therefore the meaning of the synchronization
remains the same.
\end{claim}

\subsection{Phase space}

In \textcite{GS10,algebra2011} we described execution in terms of
basic blocks and a control flow graph (CFG), a standard concept in
compiler theory (see \textcite{Aho85,AptOld97}) A state  parallel
execution can be identified as a list of blocks that are executing
concurrently at a given time $t$, and possible successor states are
given by the CFG. A synchronization is defined by a set of events
each occurring at a different block and having a particular time relation
with events at other blocks. For $N$ processes, this gave us a phase
space as an $N$sided hypercube, with each side labeled with code
block numbers. A state would be an ordered $N$ tuple of block labels,
and as such would uniquely identify a point in the phase space hypercube.

To relate Algebra of Synchronization hi described here and in~\textcite{GS10,algebra2011}
to the phase space hypercube, we note that numeric relations restrict
the set of possible states. For example, $A>B$ describes sets such
that every number in $A$ is greater than every number in $B$. If
the numbers $A$ and $B$ represent block numbers executed concurrently
by two processes, this relation blocks states $a_{i}\in A,b_{j}\in B$
with $a_{i}$ earlier than $b_{j}$. From the control flow graph we
can predict successor states to both $a_{i}$ and $b_{i}$ that are
in blocked regions of the phase space. We can see the effect of forced
waits in the execution graphs in \textcite{GS10,algebra2011}, as
dense clumps of states resulting from processes blocked from advancing
by a synchronization condition.

We can see what happens to each specific event in a synchronization
from the boundedness condition of events on the diagonal of the synchronization
matrix (subsection \ref{sec:Calculation-of-boundedness}) . In continuing
work, we will define how each boundedness condition will reduce the
accessible states from each event on the diagonal and how this will
affect the phase space. This in turn will be used to predict the effect
on phase space hypervolume, entropy, time and work costs of synchronization,
and verify predictions experimentally.

\section{Open problems}

\subsection{Equivalence}

We claim that two synchronization matrices are equivalent IFF they
have the same closure - do we need a separate theorem to prove this?
If the closure algorithm is correct, then the proof is easy, because
closure gets all the implied synchronizations. We may need to consider
semaphores as implemented in combination with queues, however ( and
possibly other cases in which a relation changes with time)

An implication of EVENTSWAP, is that any permutation of events in
the synchronization matrix is equivalent to every other permutation.
This property should extend to every matrix that has the same closure,
since EVENTSWAP preserves the boundedness of events being swapped. 

\subsection{Separability}

When can we say that a set of individual binary synchronizations constitute
a collective synchronization? Suppose we have a set of events that
occur in a loop which runs in parallel on multiple processes. Further
suppose there are multiple synchronizations involving different processes
in the loop. We could always represent every relation in the loop
with a single synchronization matrix, but this could hide insight
about what is logically happening and what is taking more time in
execution. Understanding what the code is supposed to do, we may be
able to divide synchronizations into different sets that are logically
independent, so they could be representable by different synchronization
matrices, and possibly abstracted into collective functions..

We do not know if an algorithm exists that would allow us to identify
sets of related synchronizations.

\subsection{Optimization}

It seems (\textcite{GS16}) that the major time cost of synchronization
is in the synchronization waits and entropy reduction, rather than
in the signals or messages required to implement a synchronization.
Nevertheless we would like to know what is the simplest synchronization
code that would have a given effect on execution. We claim that the
meaning of synchronization is given by the closure of the synchronization
matrix, and we also know that different code can lead to the same
closure (due to implied synchronization). 

Therefore: given a the closure of a particular synchronization matrix,
what is the simplest matrix that produces the same closure? 

An allied question: what do we mean by simplest? Smallest number of
binary synchronizations required is the simplest, but do we need to
narrow down to something like smallest number of atoms used?

\printbibliography

\end{document}